\documentclass[twocolumn,showpacs,pre,amsmath,amssymb,floatfix]{revtex4}

\usepackage{graphicx}% Include figure files
\usepackage{dcolumn} % Align table columns on decimal point
\usepackage{bm}      % bold math
\usepackage{latexsym}
\usepackage{color}

\def\add#1{#1}                             % addition
\def\ad#1{#1}                              % addition
\newcommand{\vomega}{\mbox{\boldmath $ \omega $}}
\newcommand{\vmu}{\mbox{\boldmath $ \mu $}}
\begin{document}

\title{Magnetohydrodynamic activity inside a sphere}

\author{Pablo D. Mininni$^1$ and David C. Montgomery$^2$}
\affiliation{$^1$ National Center for Atmospheric 
Research, P.O. Box 3000, Boulder, Colorado 80307}
\affiliation{$^2$ Dept. of Physics and Astronomy,
Dartmouth College, Hanover, NH 03755} 

\date{\today}

\begin{abstract}
\add{We present a computational method to solve the magnetohydrodynamic 
equations in spherical geometry. The technique is fully nonlinear and 
wholly spectral, and uses an expansion basis that is adapted to the 
geometry: Chandrasekhar-Kendall vector eigenfunctions of the curl. The 
resulting lower spatial resolution is somewhat offset by being able to 
build all the boundary conditions into each of the orthogonal expansion 
functions and by the disappearance of any difficulties caused by 
singularities at the center of the sphere. The results reported here 
are for mechanically and magnetically isolated spheres, although different 
boundary conditions could be studied by adapting the same method. The 
intent is to be able to study the nonlinear dynamical evolution of those 
aspects that are peculiar to the spherical geometry at only moderate 
Reynolds numbers. The code is parallelized, and will preserve to high 
accuracy the ideal magnetohydrodynamic (MHD) invariants of the system 
(global energy, magnetic helicity, cross helicity). Examples of results 
for selective decay and mechanically-driven dynamo simulations are 
discussed. In the dynamo cases, spontaneous flips of the dipole 
orientation are observed.}
\end{abstract}

\pacs{47.11.-j; 47.11.Kb; 91.25.Cw; 95.30.Qd}
\maketitle

\section{\label{sec:intro}INTRODUCTION}

Magnetohydrodynamic ``dynamo'' processes are those in which the motions 
of an electrically conducting fluid amplify and maintain a finite magnetic 
field, starting from an arbitrarily small one. They have long been of 
interest for geophysics and astrophysics 
\cite{Moffatt,Glatzmaier95,Dikpati99,Kono02,Nandy02,Mininni04}, and have 
more recently become of interest with regard to laboratory attempts to 
generate dynamo magnetic fields in liquid metals 
\cite{Gailitis01,Steiglitz01,Noguchi02,Petrelis03,Sisan03,Spence05}.
The relevant theoretical and computational literature is vast, and 
extensive reviews have recently been given (see e.g. Refs. 
\cite{Roberts01,Kono02,Brandenburg05}).

In our own work, we have lately been studying dynamo processes numerically, 
using turbulent three-dimensional magnetohydrodynamic (hereafter, ``MHD'') 
codes of the familiar Orszag-Patterson pseudospectral variety 
\cite{Ponty05,Mininni05a,Mininni05b,Alexakis05,Mininni05c}. Such codes 
treat homogeneous turbulence efficiently, but are mainly useful in 
situations involving spatially periodic boundary conditions. Particularly 
for the case of planetary dynamos and laboratory experiments, restrictions 
to periodic boundary conditions are a severe limitation. Essential 
ingredients, such as rotation, global angular momentum, and the 
interfaces between conducting and non-conducting regions are not 
readily included. It is to be expected that all of these play a 
role in the physical situations of interest, and give rise to 
qualitatively new physical processes not accessible with periodic 
boundaries.

This present paper represents our attempt to begin a study of these 
processes by introducing, and displaying some results from, a 
computational method that is adapted to the geometry of isolated 
spheres. Our goal is not to reach realistic geophysical parameter 
regimes (out of the question for the foreseeable future, in any case), 
but rather to isolate and study those new physical processes that appear 
in this geometrically more realistic setting. \add{Our results are not 
to be compared with the ambitious geo-dynamo computation of Glatzmaier 
and Roberts (e.g., Refs. [2,21]) who include far more physical effects} 
\ad{and degrees of freedom} \add{than we have at this point. We will 
consider it not to be a limitation on our work if we are unable to reach 
realistic Reynolds numbers, so long as the spectra of the dynamo 
processes we do identify are well-resolved by the number of expansion 
functions we are able to retain. We think of these efforts as studies 
of dynamo behaviors exhibited by the MHD equations in spherical geometry 
without regard to their present applicability to geophysical or laboratory 
situations. It appears not to be necessary to reach presently-existing 
ranges of Reynolds numbers or magnetic Prandtl numbers in order to see 
interesting spontaneously-generated magnetic field behavior.}

\add{The numerical technique is entirely spectral, using an expansion 
basis that is specifically adapted to spherical geometry: the 
Chandrasekhar-Kendall (hereafter, ``C-K'') vector eigenfunctions of 
the curl \cite{Chandrasekhar57,Turner83,Yoshida91,Yoshida92,Cantarella00} 
(by construction, C-K functions are also eigenfunctions of the Helmholtz 
wave equation; see e.g. \cite{Mueller}). These functions form a complete 
orthogonal set under the boundary conditions we will choose. They can 
be normalized. Their completeness has been shown by Yoshida for the 
cylinder \cite{Yoshida92} and by Cantarella et al \cite{Cantarella00} for 
the sphere. The cylindrical version was used some years ago for studying 
processes believed to be involved in the disruptions of fusion confinement 
devices \cite{Montgomery78,Shan91,Shan93a,Shan93b,Shan94}. Their advantages 
lie in their natural geometrical relation to the specific geometries in 
which they are employed (all the boundary conditions can be built into 
each expansion function) and in certain desirable numerical properties, 
which include the following. The MHD equations involve several solenoidal 
fields of which several curls are taken. Taking the curl of a C-K function 
simply gives the function back again, times a multiplicative constant. 
This means that no numerical spatial differentiations are required, with 
their attendant complication of the expressions and loss of accuracy. The 
code preserves the ideal MHD invariants very accurately over long times 
(total energy, magnetic helicity, cross helicity), which is one of the 
few accuracy checks available to a strongly nonlinear code for which 
non-trivial analytical solutions are scarce. In addition, the ideal 
invariants are readily exhibited as simple algebraic quadratic sums 
involving the expansion coefficients, so no numerical integrations are 
required to evaluate them. Going along with these advantages is a severe 
disadvantage associated with the lack of a fast transform for the 
spherical Bessel functions involved, which means that the nonlinear terms 
lead to convolution sums which grow rapidly with the number of expansion 
functions retained in the Galerkin approximation. This limits us to 
mechanical and magnetic Reynolds numbers of the order of a very few 
thousand, and makes it unlikely that we can ever reach planetary parameters 
without modeling the small scales of the fluid motions.}

\add{But the absence of any coordinate singularities (e.g., $r=0$ in 
spherical polar coordinates) to worry about anywhere is a considerable 
advantage, as we noted previously in solving the two-dimensional 
Navier-Stokes equation inside a circle with non-ideal boundaries 
\cite{Li96a,Li97}. The code also can be readily parallelized and there 
are no potential aliasing problems.}

\add{The situation we wish to study is that of an electrically conducting 
fluid inside a rigid spherical boundary that isolates the fluid, 
mechanically and magnetically, from everything outside it. The boundary 
is regarded as a spherical shell that is perfectly conducting (so 
magnetic fields cannot penetrate the region outside) and is coated on 
the inside with a thin layer of insulating dielectric so that the 
electrical current density cannot penetrate the shell. The shell is 
regarded as mechanically impenetrable, so that the normal component of 
the fluid velocity vanishes there. We do not employ no-slip boundary 
conditions on the velocity field, but rather choose the vanishing of the 
normal component of the vorticity at the spherical boundary as the second 
boundary condition on the velocity field. This condition is implied by, 
but does not imply, no-slip boundary conditions on the velocity field, 
and thereby avoids certain mathematical procedures that sometimes attend 
the attempts at imposing no-slip boundary conditions.} \ad{These four 
boundary conditions for the fields in the external walls, plus regularity 
of each component of the velocity and magnetic fields at the origin, might 
be thought of as a total of ten boundary conditions for the system. The 
above mentioned boundary condition for the vorticity in the wall is easy 
to implement in the present formulation. Other boundary conditions (e.g. 
no-slip boundary conditions) can be studied with our method using a 
penalty method close to the walls (as done for example in Ref. 
\cite{Shan94}) and will be considered in the future. However, we want 
to point out that the present election of boundary conditions also 
avoids some problems present in hydrodynamic incompressible flows 
when no-slip conditions are imposed (see e.g. \cite{Kress00,Gallavotti} 
for a detailed discussion). Since this topic is beyond the aim of our 
present study, we will use in the following these simple boundary 
conditions and leave the study of other options for future work.}

\add{An outline of the paper follows. Section \ref{sec:equations} 
introduces the expansion basis and shows how the nonlinear partial 
differential equations can be reduced to a set of ordinary differential 
equations, first order in the time, for the expansion coefficients. We 
include mention of other possible boundary conditions that it is 
intended to introduce in the future, then settle on the one just 
described for these runs. The numerical method is discussed in Section 
\ref{sec:numerical}. Sections \ref{sec:application1} and 
\ref{sec:application2} display our first applications of the code to 
some simple MHD problems that are considered interesting and that 
are peculiar to spherical geometry. We simultaneously explore 
the possibilities for some relative new flow visualization diagnostics 
when exhibiting our results \cite{vapor}; these are described when 
they appear. Section VI summarizes the results and discusses possible 
future applications of the method.}

\section{\label{sec:equations}THE SPECTRAL DECOMPOSITION}

C-K functions are constructed from a solution of the scalar Helmholtz 
equation:
\begin{equation}
\left( \nabla^2 + \lambda^2 \right) \psi = 0 ,
\label{eq:helmholtz}
\end{equation}
where we are referring to spherical polar coordinates $(r,\theta,\phi)$ 
and $\lambda$ is an eigenvalue that will eventually be determined by 
boundary conditions. Vector eigenfunctions of the curl appropriate to 
spherical geometry may be constructed according to the \add{following} 
recipe \add{from each solution to Eq. (\ref{eq:helmholtz})}:
\begin{equation}
{\bf J} = \lambda \nabla \times {\bf r} \psi + \nabla \times \left( 
    \nabla \times {\bf r} \psi \right) .
\end{equation}
From this, it may readily be verified that 
$\nabla \times {\bf J} = \lambda  {\bf J}$. Thus any single ${\bf J}$ 
is a ``force-free'' or ``Bernoulli'' field, though the sum of two or 
more of them is not. The relevant scalar $\psi$ is
\begin{equation}
\psi_{qlm} = C_{ql} \, j_l(|\lambda_{ql}| r) Y_{lm} (\theta,\phi) .
\label{eq:psi}
\end{equation}
Here, $C_{ql}$ is a normalization constant, $j_l$ is a spherical Bessel 
function of order $l$, and $Y_{lm}$ is the normalized spherical harmonic 
expressed in terms of the polar angle $\theta$ and the azimuthal 
(longitudinal) angle $\phi$. The number $C_{ql}$ is chosen to make 
the volume integral of ${\bf J}_{qlm} \cdot {\bf J}_{qlm}^*$ over the 
computational domain equal to unity (the asterisk denotes complex 
conjugate).

The integer $l$ is $1,2,3,\dots$ and $m$ runs in integer steps from $-l$ 
to $l$. An infinite sequence of values of $\lambda_{ql}$, labeled by 
$q=1,2,3,\dots$ and $q=-1,-2, -3,\dots$ \add{will be determined by a 
radial boundary condition momentarily to be invoked} on ${\bf J}$ for 
a given $l$.

\add{The boundary conditions that will be invoked, consistently with the 
physical description of the region and its boundaries that were given in 
Sec. \ref{sec:intro} refer to the magnetic field ${\bf B}$ and the 
velocity field ${\bf v}$, both solenoidal, and their curls. The curl of 
${\bf v}$ is $\vomega$, the vorticity, and the curl of ${\bf B}$ is 
${\bf j}$, the electric current density, in the dimensionless units that 
remain to be described.  We shall demand that the normal (radial) 
components of all four vector fields vanish at the boundary $r=R$. Since 
these four fields are to be expanded in the ${\bf J}$'s, making the 
radial component of each ${\bf J}$ vanish at $r=R$ will guarantee that 
the boundary conditions will be satisfied for any superposition of them. 
This vanishing of the normal components of the ${\bf J}$ determines the 
allowed values of $\lambda_{ql}$, positive and negative, by the locations 
of the zeros of the spherical Bessel functions. Then the explicit 
expression for the normalization constant $C_{ql}$ is}
\begin{equation}
C_{ql} =  \left|\lambda_{ql} \, j_{l+1}(|\lambda_{ql}|R) \right|^{-1} 
    \left[ l(l+1) R^3 \right]^{-1/2} .
\label{eq:normalization}
\end{equation}
Positive or negative $q$ is to be associated with positive or negative 
$\lambda_{ql}$ according to $\lambda_{-q,l} = - \lambda_{ql}$. For a 
variety of radial boundary conditions, \add{including the one just chosen,} 
the ${\bf J}_{qlm}$ functions corresponding to differing indices $q$, $l$, 
or $m$ can be shown to be orthogonal. Boundary conditions at two different 
radii can be imposed if spherical Neumann functions are also permitted in 
$\psi$.

\add{Given} the completeness of the C-K functions so defined, they are then 
appropriate for expanding various vector fields of incompressible MHD: 
the magnetic field, fluid velocity, vorticity, electric current density, 
and vector potential in the Coulomb gauge. \add{These are considered to 
be the necessary set of quantities to be expanded when studying spherical 
MHD and dynamos in the class of problems studied here.} The ones of 
slowest spatial variation (smaller $|q|$, low $l$ and $|m|$) contain the 
dipole and low-order multipole components of the fields and provide a 
natural ordering of the spectral contributions from various spatial scales, 
\add{starting with those of the largest scales.}

The MHD equations to be solved are, in familiar dimensionless 
(``Alfv\'enic'') units, \add{common in MHD turbulence theory,} 
an equation of motion for the fluid velocity 
${\bf v}$,
\begin{equation}
\frac{\partial {\bf v}}{\partial t} = {\bf v} \times \vomega + 
    {\bf j} \times {\bf B} - \nabla \left({\mathcal P} + \frac{v^2}{2} 
    \right) + \nu \nabla^2 {\bf v} + {\bf f} , 
\label{eq:momentum}
\end{equation}
and the induction equation for advancing the magnetic field ${\bf B}$,
\begin{equation}
\frac{\partial {\bf B}}{\partial t} = \nabla \times \left( {\bf v} 
    \times {\bf B} \right) + \eta \nabla^2 {\bf B} ,
\label{eq:induction}
\end{equation}

\add{In these units, ${\bf v}$ may be considered to be normalized by 
the rms value of the velocity field ($U$, say), and the magnetic field 
in units that, using the square root of the mass density, converts a magnetic 
field to an Alfv\'en speed based upon the same rms velocity. In Eqs. 
(\ref{eq:momentum}) and (\ref{eq:induction}), lengths are measured in 
units of the spherical radius $R$ (the radius $R$ will be 1 in the 
dimensionless units) and times in units of $R/U$.} ${\mathcal P}$ is 
the dimensionless ratio of pressure to mass density, with the mass 
density assumed to be spatially uniform. A forcing function ${\bf f}$ 
has been written on the right hand side of Eq. (\ref{eq:momentum}) to 
represent any externally applied mechanical force that can be chosen 
to mimic a variety of physical effects, a common convention in 
dynamo-motivated computations \add{in periodic boundary conditions}. 
Both ${\bf B}$ and ${\bf v}$ and their curls are solenoidal. \ad{Note 
that in the incompressible case, nonlinear coupling between modes is 
provided by the nonlinearities in both the equation of motion and the 
induction equation. Here, we are keeping all of these. In the context 
of planetary dynamos, other physical effects sometimes justify dropping 
the advection term in the equation of motion. Then other numerical 
methods (see e.g. Ref. \cite{Ivers03}) for linearised MHD can be used.}

An evolution equation for vorticity can be obtained by taking the curl of 
Eq. (\ref{eq:momentum}):
\begin{equation}
\frac{\partial \vomega}{\partial t} = \nabla \times \left( 
    {\bf v} \times \vomega \right) +  \nabla \times \left( 
    {\bf j} \times {\bf B} \right) + \nu \nabla^2 \vomega + 
    \nabla \times{\bf f} .
\label{eq:vorticity}
\end{equation}

\add{Given the} appropriate boundary conditions, the vorticity $\vomega$ 
can be used to determine the velocity ${\bf v}$. The dimensionless viscosity 
$\nu$ and magnetic diffusivity $\eta$ can be interpreted, respectively, as 
reciprocals of kinetic and magnetic Reynolds numbers based on $U$, $R$, and 
the laboratory (dimensional) values of kinematic viscosity and magnetic 
diffusivity.

The numerical scheme is to solve Eqs. (\ref{eq:momentum}) or 
(\ref{eq:vorticity}) and (\ref{eq:induction}) by representing ${\bf v}$ 
and ${\bf B}$ as the Galerkin expansions,
\begin{eqnarray}
{\bf v}({\bf r},t) &=& \sum_{qlm} \frac{\xi^v_{qlm}(t)}{\lambda_{ql}}
    {\bf J}_{qlm}({\bf r}) \;\;\;\; (r \le R) \label{eq:expanv} \\
{\bf B}({\bf r},t) &=& \sum_{qlm} \xi^B_{qlm}(t) \, {\bf J}_{qlm}({\bf r}) .
\label{eq:expanb}
\end{eqnarray}
Here, the unknowns are the \add{now} complex, time-dependent expansion 
coefficients $\xi^v_{qlm}$ and $\xi^B_{qlm}$. The vorticity and current 
density are given by the same series, each term multiplied by the 
appropriate value of $\lambda_{ql}$.

Next, we may substitute Eqs. (\ref{eq:expanv}) and (\ref{eq:expanb}) 
into Eqs. (\ref{eq:induction}) and (\ref{eq:vorticity}), say, and take 
inner products one at a time with the functions ${\bf J}_{q'l'm'}$. 
Using their orthonormality, the dynamical equations become
\begin{eqnarray}
\frac{\partial \xi^v_i}{\partial t} &=& \sum_{j,k} \left( A^i_{jk} 
    \xi^v_j \xi^v_k + B^i_{jk} \xi^B_j \xi^B_k \right) - \nu \lambda_i^2 
    \xi^v_i + \xi^f_i 
\label{eq:CK1} \\
\frac{\partial \xi^B_i}{\partial t} &=& \sum_{j,k} C^i_{jk} \xi^v_j \xi^B_k
    - \eta \lambda_i^2 \xi^B_i .
\label{eq:CK2}
\end{eqnarray}

The indices $i,j,k$ are regarded as shorthand; each one of them represents 
the triple of numbers $q,l,m$ necessary to identify a single member of the 
family of the C-K functions ${\bf J}_{qlm}$. The sum is over all the values 
retained in the Galerkin expansion.

The nonlinearities, both in the original partial differential equations and 
in the ordinary differential equations (\ref{eq:CK1}) and (\ref{eq:CK2}), 
are quadratic. The kinematic coupling coefficients (which do not contain 
the $\xi$) $A^i_{jk}$, $B^i_{jk}$, and $C^i_{jk}$, are numerical integrals 
of considerable complexity. Their evaluation and storage as a table is one 
of the most demanding parts of the computation, and some features of that 
evaluation are described in Section \ref{sec:numerical}. On the right hand 
side of Eq. (\ref{eq:CK1}), we have represented the forcing function 
${\bf f}$, if any, by its expansion over the retained C-K functions. The 
coefficients $\xi^f$, if non-zero, are to be regarded as known, possibly 
time-dependent functions that may excite the velocity field, and 
may stand for various mechanical processes.

The coupling coefficients in Eqs. (\ref{eq:CK1}) and (\ref{eq:CK2}) are 
reducible to
\begin{equation}
A^i_{jk} = C^i_{jk} = \frac{\lambda_i}{\lambda_j} I^i_{jk} \, , 
\;\;\;\;\;\; B^i_{jk} = \lambda_i \lambda_j I^i_{jk} \, ,
\end{equation}
where
\begin{equation}
I^i_{jk} = \int {\bf J}^*_i \cdot \left( {\bf J}_j \times {\bf J}_k 
    \right) {\textrm d}^3 x .
\label{eq:integral}
\end{equation}

\section{\label{sec:numerical}NUMERICAL METHOD}

\begin{figure}
\includegraphics[width=9cm]{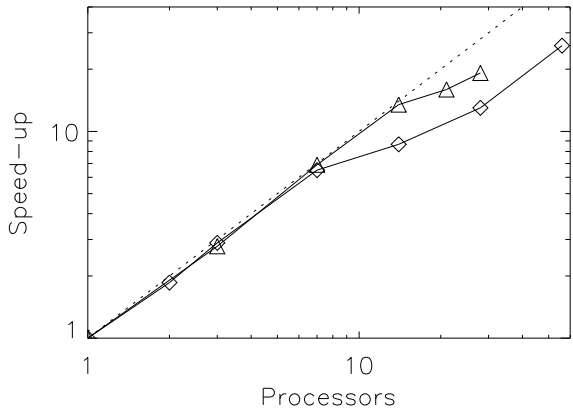}
\caption{Speed-up of the code in two Linux clusters in parallel 
    simulations with $\max\{q\} = \max\{l\} = 7$. The dotted line 
    indicates the ideal scaling.}
\label{fig:speedup}
\end{figure}

Equations (\ref{eq:CK1}) and (\ref{eq:CK2}) are solved numerically with 
double precision in a sphere of radius $R=1$. The expansion coefficients 
$\xi$ are in general complex, and since the fields are real, the 
coefficients satisfy the condition $\xi_{q,l,-m} = (-1)^m \xi^*_{qlm}$. 
As a result, only the coefficients for non-negative values of $m$ are 
stored and evolved in time.

Before the simulation is started, for a given resolution in $q$ and $l$ 
(and all possible values of $m$) all required values of the normalization 
coefficients $C_{ql}$ are computed using Eq. (\ref{eq:normalization}) 
and stored. The values of $\lambda_{ql}$ are computed numerically as 
the roots of the spherical Bessel functions using a combination of 
bisection and Newton-Raphson \cite{Press}. Finally, the coupling 
coefficients $I^i_{jk}$ are computed and stored.

The coupling coefficients are complex, and from Eq. (\ref{eq:integral}) 
satisfy the relation $I^i_{jk} = - I^i_{kj}$. The integral in Eq. 
(\ref{eq:integral}) is separable in spherical coordinates. In the $\phi$ 
direction, the integral reduces to the condition $m_k = m_i-m_j$; in 
any other case the coupling coefficients $I^i_{jk}$ are zero. The 
radial integral reduces to seven integrals involving three spherical 
Bessel functions or their derivatives, and the integral in the polar 
angle reduces to seven integrals on three Legendre functions and their 
derivatives. Radial integrals are computed numerically with high 
precision using Gauss-Legendre quadratures, while integrals in $\theta$ 
are computed using Gauss-Jacobi quadratures \cite{Press}. Due to 
symmetry properties of the Legendre functions, all coupling 
coefficients with $l_i+l_j+l_k+m_i+m_j+m_k$ even are purely real, 
while the remaining coefficients are purely imaginary, another 
property that can be used to save memory.

Once tables containing all these values are stored, the evolution of 
the system reduces to solving the set of ordinary differential 
equations defined by Eqs. (\ref{eq:CK1}) and (\ref{eq:CK2}). These 
equations are evolved using a Runge-Kutta method of fourth order 
\cite{Canuto}. The MHD equations have three quadratic ideal invariants: 
the total energy $E$, the magnetic helicity $H$, and the cross helicity 
$K$. In spectral space the invariants can be computed as
\begin{eqnarray}
E &=& \frac{1}{2} \sum_i \left( \frac{|\xi^v_i|^2}{\lambda_i^2} + 
    | \xi^B_i|^2 \right) , \\
H &=& \frac{1}{2} \sum_i \frac{|\xi^B_i|^2}{\lambda_i} , \\
K &=& \frac{1}{2} \sum_i \frac{\xi^v_i \xi^{B*}_i}{\lambda_i} .
\end{eqnarray}
The conservation of these quantities up to the numerical precision 
serves as a test of the code and have been verified in simulations 
with $\nu=\eta=0$. In a simulation with $\max\{q\} = \max\{l\} = 5$, 
the invariants were conserved up to the sixth decimal place after 
200 turnover times. The turnover time is defined as $T=R/U$.

The system is evolved entirely in spectral space and all global 
quantities are also computed spectrally. To obtain representations 
of the fields in configuration space, Eqs. (\ref{eq:expanv}) and 
(\ref{eq:expanb}) are used.

The reciprocal of the smallest $|\lambda|$ may be identified with the 
largest length scale in the dynamics allowed by the boundary conditions, 
and the reciprocal of the largest $|\lambda|$ retained may be considered 
to be the smallest resolvable spatial scale. In a typical computation 
(see e.g. Section \ref{sec:application2}), these numbers may be 
$\approx 4.59$ and $\approx 41.3$, respectively. The minimum and maximum 
wavenumbers in our previous 3D periodic dynamo computations (e.g. 
in a $256^3$ dealiased simulation using the $2/3$-rule \cite{Canuto}) 
have typically been 1 and 85 in dimensionless units, by comparison.
\add{The fact that the maximum to minimum ratio of allowed wavenumbers 
is more than $9$ times greater for the $256^3$ code shows that far 
less spatial resolution appears to be available in the C-K code as 
presently run. If ``degrees of freedom'' are taken as a formal measure 
of the resolution, then its $1.6\times10^4$ makes the C-K code roughly 
comparable with the $7.8\times10^4$ available to a de-aliased periodic 
3D code that is of resolution $64^3$. That is not, however, the whole 
story, since very many Fourier coefficients would be needed to represent 
accurately any of the C-K functions which are, individually, all 
consistent global physical states of the system obeying all the 
boundary conditions. We find that there are many situations that can be 
computed with the C-K code that are nontrivial, highly nonlinear, and 
that are well-resolved as indicated by the presence of an unmistakable 
dissipation range for their $\lambda$-spectra; these include situations 
exhibiting a variety of dynamo behavior}.

Besides the quadratic ideal invariants and the reconstruction of 
the field components, two vector quantities will be of interest. The 
angular momentum ${\bf L}$ is defined as
\begin{equation}
{\bf L} = \int {\bf r} \times {\bf v} \, {\textrm d}^3 x \, ,
\end{equation}
where a unity mass density is assumed, and the magnetic dipole moment 
$\vmu$ is given by
\begin{equation}
\vmu = \frac{1}{2} \int {\bf r} \times {\bf j} \, {\textrm d}^3 x \, .
\end{equation}
In terms of the C-K functions, these two quantities become
{\setlength\arraycolsep{2pt}
\begin{eqnarray}
{\bf L} &=& 4 R^3 \sqrt{\frac{\pi}{3}} \sum_q C_{q,1} 
    \frac{j_1'(|\lambda_{q,1}| R)}{|\lambda_{q,1}|} \left[ 
    -\xi^v_{q,1,0} \hat{z} + \right. \nonumber \\
&& {} \left. + \sqrt{2} {\textrm Re} \left(\xi^v_{q,1,1}\right) \hat{x} - 
    \sqrt{2} {\textrm Im} \left(\xi^v_{q,1,1}\right) \hat{y} \right] , \\
\vmu &=& 2 R^3 \sqrt{\frac{\pi}{3}} \sum_q C_{q,1} |\lambda_{q,1}| 
    j_1'(|\lambda_{q,1}| R) \left[ -\xi^B_{q,1,0} \hat{z} + \right. 
    \nonumber \\
&& {} \left. + \sqrt{2} {\textrm Re} \left(\xi^B_{q,1,1}\right) \hat{x} - 
    \sqrt{2} {\textrm Im} \left(\xi^B_{q,1,1}\right) \hat{y} \right] .
\end{eqnarray}}

As previously mentioned, in the code a sphere with $R=1$ is used. 
Note that only modes with $l=1$ and $m=0, \pm1$ give a contribution to 
the angular momentum and the dipole moment. With the boundary conditions 
considered in this work, the angular momentum is not a conserved quantity 
unless $\nu = 0$. Other choices of the boundary conditions can lead to a 
conservation of the angular momentum even in the non-ideal case. Those 
boundary conditions apply to the case when the sphere of magnetofluid is 
isolated from torques, but we will defer consideration of that situation 
to a later paper.

The code is written in Fortran 90 and parallelized using MPI. Since 
most of the computing time is spent in the sums in Eqs. (\ref{eq:CK1}) 
and (\ref{eq:CK2}), the parallelization is done as follows. Each 
processor contains a complete copy of the expansion coefficients 
$\xi$, but only a portion of the coupling coefficients $I^i_{jk}$. 
The array $I^i_{jk}$ is distributed in $q$ if the number of processors 
is smaller or equal than $2q$, and distributed in $q,l,m$ in any other 
case. Each processor computes the sums in Eqs. (\ref{eq:CK1}) and 
(\ref{eq:CK2}) locally for the corresponding values of $q,l,m$, and 
after each iteration of the Runge-Kutta method the coefficients 
$\xi$ are synchronized between all processors. The required 
communication is minimal and the scaling of the code as the number 
of processors is increased is close to ideal.

Figure \ref{fig:speedup} shows the speed-up vs. the number of 
processors in two different Linux clusters, in a simulation using 
$\max\{q\} = \max\{l\} = 7$. The clusters differ in the network 
configuration. The speed-up is defined as the time required to do 
one time step in $N$ processors divided by the time required in 
one processor. The code shows ideal scaling up to $N\approx 2 \max\{q\}$. 
A drop is then observed and is related to the change in the parallel 
distribution of the array $I^i_{jk}$. However, after this drop a 
linear scaling is again recovered.

\section{\label{sec:application1}SELECTIVE DECAY}

\begin{figure}
\includegraphics[width=9cm]{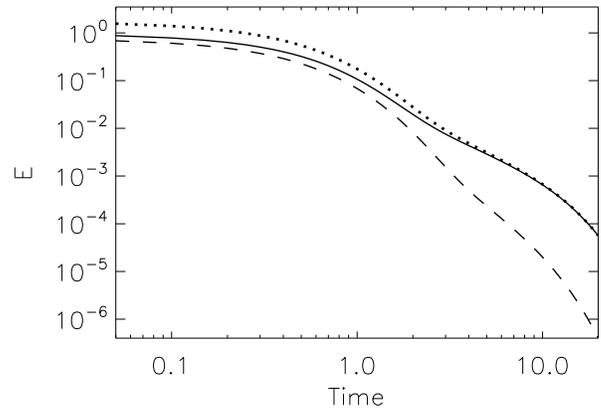}
\caption{Total energy (dotted line), magnetic energy (solid line), and 
    kinetic energy (dashed line) as a function of time in run II. At 
    late times the system is dominated by magnetic energy. \add{Since the 
    only dynamic evolution possible for ${\bf B}$ requires a velocity 
    and since the kinetic energy has essentially disappeared, the system 
    has ``frozen'' into a nearly purely magnetic state, which can only 
    slowly resistively decay hereafter.}}
\label{fig:decay_ener}
\end{figure}

\begin{figure}
\includegraphics[width=9cm]{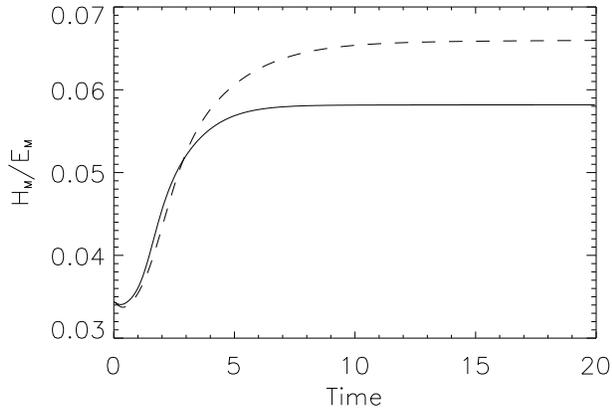}
\caption{Relative magnetic helicity as a function of time for runs I 
    (solid line) and II (dashed line). The relative helicity at late 
    times increases with the Reynolds number. Note the maximum 
    possible value for the relative helicity is 
    $\min^{-1}\{|\lambda|\} \approx 0.22$.}
\label{fig:decay_hel}
\end{figure}

\add{``Selective decay'' refers to a frequently studied MHD turbulence 
situation in which a system with initial magnetic helicity present 
evolves with a  rapid decay of total energy relative to the magnetic 
helicity. This situation has been studied previously in periodic 
boundary conditions and it is of interest to see if the behavior 
is affected by spherical geometry.} The late-time state is a quasi-steady 
configuration in which the remaining energy is nearly all magnetic 
and is condensed into the longest wavelength modes allowed by the 
boundary conditions. Selective decay has been extensively studied in 
periodic boundary conditions (see, e.g., \cite{Matthaeus80,Ting86,Kinney95}). 
In Ref. \cite{Mininni05d} it was found that given isotropic initial  
conditions in a periodic box, the final state of the magnetic field 
corresponds to an Arn'old-Beltrami-Childress (ABC) field at the largest 
possible scale with $A$, $B$, and $C$ equal. It is therefore of interest 
to test selective decay in a sphere, and to observe the geometry 
of the magnetic and velocity fields in the late-time state.

Two runs were done, the first (run I) with $\nu = \eta = 1\times 10^{-2}$, 
and the second (run II) with $\nu = \eta = 6\times 10^{-3}$. Run I was 
done with $\max \{q\} = \max \{l\} = 7$, and run II with 
$\max \{q\} = \max \{l\} = 9$. In both simulations, no external 
force was applied, and the system was allowed to evolve for a 
long time (20 initial large scale turnover times).

The non-vanishing initial expansion coefficients in both simulations are
\begin{eqnarray}
& \xi^v_{3,3,0} = \xi^v_{-3,3,0} = -u_0 , \;\;
    \xi^B_{3,3,0} = \frac{5}{3} \xi^B_{-3,3,0} = B_0 , \\
& \xi^v_{3,3,m} = \xi^v_{-3,3,m} = u_0(1+i) , \\
& \xi^B_{3,3,m} = \frac{5}{3} \xi^B_{-3,3,m} = B_0(1-i) ,
\end{eqnarray}
where $m$ runs from 1 to 3 and negative values of $m$ are given by 
$\xi_{q,l,-m} = (-1)^m \xi^*_{qlm}$. The amplitudes $u_0$ and $B_0$ were 
chosen to have initial kinetic and magnetic energies of order unity 
($u_0 = 4$ and $B_0 = 0.4$). These initial conditions correspond to 
a small cross correlation between the velocity and magnetic fields, a 
non-helical velocity field at an intermediate scale, and a 
(non-maximally) helical magnetic field at the same scale. The 
initial angular momentum is zero and remains negligible during 
the complete simulation. The kinetic and magnetic Reynolds numbers, 
based on the length $R=1$ and the initial rms velocity, are 
respectively defined as
\begin{eqnarray}
R_V & = & \frac{R U}{\nu} , \\
R_M & = & \frac{R U}{\eta} .
\end{eqnarray}
In run I $R_V = R_M \approx 98$, and in run II $R_V = R_M \approx 165$.

\begin{figure}
\includegraphics[width=9cm]{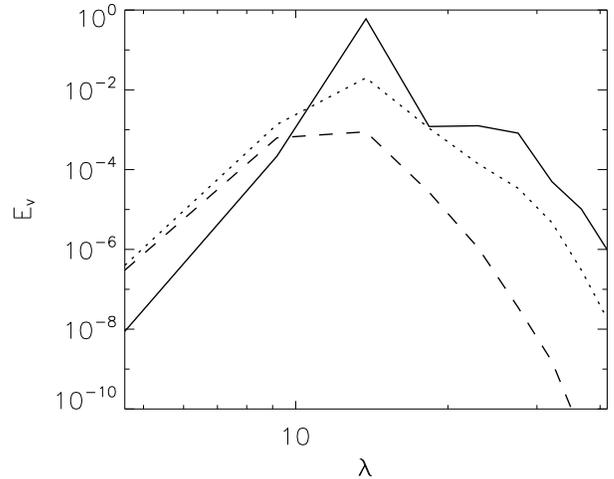}
\caption{Spectrum of kinetic energy as a function of $\lambda$ for run 
    II, at $t=0.1$ (solid line), $t=1.5$ (dotted line) and $t=3$ (dashed 
    line). \add{Note the appearance of a clear dissipation range.}}
\label{fig:decay_ksp}
\end{figure}

\begin{figure}
\includegraphics[width=9cm]{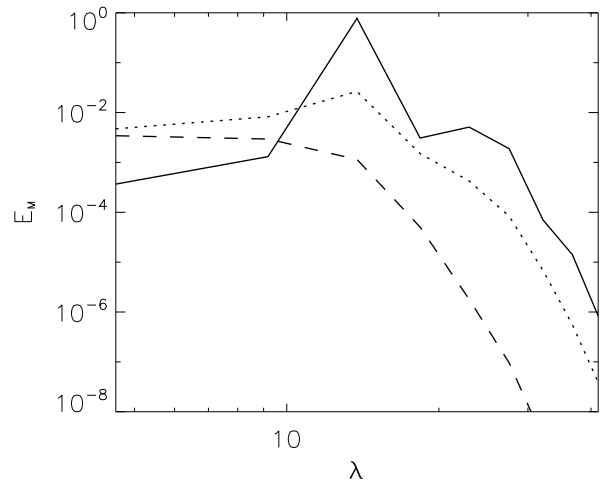}
\caption{Spectrum of magnetic energy as a function of $\lambda$ for run 
    II, at $t=0.1$ (solid line), $t=1.5$ (dotted line) and $t=3$ (dashed 
    line). \add{A magnetic dissipation range becomes clearly visible.}}
\label{fig:decay_msp}
\end{figure}

Figure \ref{fig:decay_ener} shows the time history of the kinetic, 
magnetic, and total energies in run II. At late times, the kinetic 
energy is negligible and the system is dominated by magnetic energy. 
The magnetic helicity decays slowly compared to the energy, as 
indicated by the evolution of the relative helicity $H_M/E_M$ (Fig. 
\ref{fig:decay_hel}). As time evolves, the ratio $H_M/E_M$ increases 
until a steady state is reached. The final state reached by the 
system is not a ``Taylor state'', a state of maximum possible helicity 
for the given energy (the maximum possible value for the relative helicity 
is $\min^{-1}\{|\lambda|\} \approx 0.22$). In run I, this is because 
after $t\approx 5$ most of the kinetic energy has decayed, and as a 
result the spectral exchange between modes stopped. In run II the decay 
of the kinetic energy takes place at a later time, and as a result the 
final value of the relative helicity is larger than in run I. Larger final 
values of the relative helicity can be expected if the Reynolds numbers 
are further increased.

Since the C-K functions are eigenfunctions of the curl with 
eigenvalue $\lambda$, the Laplacian operators in the diffusion terms 
in Eqs. (\ref{eq:momentum}-\ref{eq:vorticity}) are proportional to 
$\lambda^2$. As a result, $|\lambda|$ plays in this case the role 
of the wavenumber $k$ in the Fourier base. To define the energy spectrum, 
we linearly bin the spectral space in shells of constant $|\lambda|$ and 
sum the power of all the coefficients in each shell, in analogy with 
the usual procedure in Fourier-based spectral methods. Figures 
\ref{fig:decay_ksp} and \ref{fig:decay_msp} show respectively the 
resulting kinetic and magnetic energy spectrum as a function of 
$|\lambda|$ in run II at three different times.

\begin{figure}
\includegraphics[width=9cm]{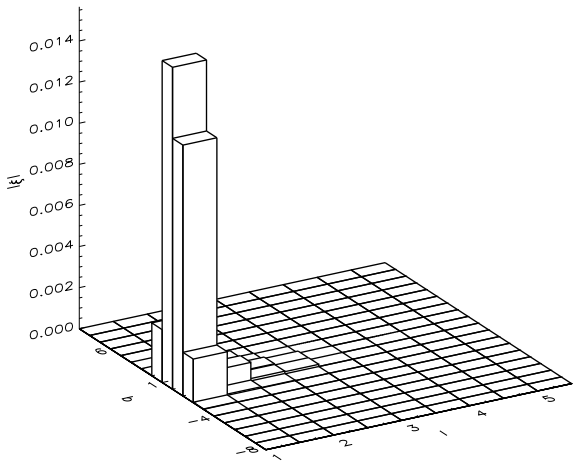}
\caption{Amplitude of the coefficients $\xi^B$ in run II at $t=10$ 
    as a function of $q$ and $l$ (summed over all values of $m$). Most 
    of the magnetic energy is at the largest available scale, and an 
    asymmetry is observed between positive and negative values of $q$.}
\label{fig:decay_xi}
\end{figure}

At early times ($t=0.1$), the signature of the initial conditions 
in the spectrum can be easily recognized. Both spectra peak at 
$|\lambda| \approx 14$, corresponding to the non-vanishing initial 
perturbation at $q=3$ and $l=3$. As time evolves, the amplitude of the 
kinetic energy spectrum decays but the position of the peak remains 
approximately constant. On the other hand, the peak in the magnetic 
energy spectrum moves to smaller values of $|\lambda|$, corresponding 
to larger scales. At late times the system is dominated by magnetic 
energy, and most of it is concentrated in the largest available scale 
in the domain.

\begin{figure*}
\includegraphics[width=14cm]{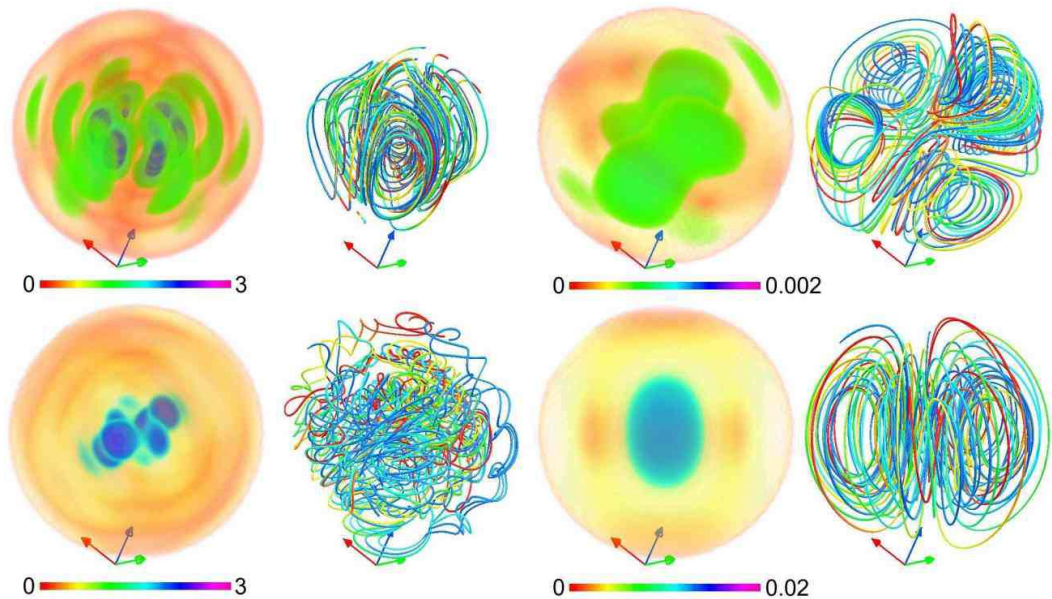}
\caption{Above: kinetic energy density and velocity field lines in run 
    II, at $t=1.5$ (left) and at $t=15$ (right). Below: magnetic energy 
    density and magnetic field lines in the same run, at $t=1.5$ (left) 
    and at $t=15$ (right). \ad{For convenience, intensity and field lines 
    are always shown in pairs, with intensity on the left and field lines 
    on the right.}}
\label{fig:decay3D}
\end{figure*}

Based on the definition of the energy spectrum, we can also 
introduce an energy-containing lengthscale as
\begin{equation}
\ell = \frac{R \, \min\{|\lambda|\}}{E} \int{E(|\lambda|) |\lambda|^{-1}
    \, {\textrm d}|\lambda|} ,
\end{equation}
and a Taylor lengthscale
\begin{equation}
\lambda_T = R \, \min\{|\lambda|\} \left[ E \bigg/ \int{E(|\lambda|) 
    |\lambda|^2 \, {\textrm d}|\lambda|} \right]^{1/2} ,
\end{equation}
where $E$ is the energy, and $E(|\lambda|)$ is the energy spectrum as 
a function of $|\lambda|$ (the sums are represented symbolically as 
integrals). Using the kinetic and magnetic energies, these characteristic 
lengths can be computed for the velocity and magnetic fields. In both 
runs, at $t=0$ $\ell \approx \lambda_T \approx 0.33$ for both fields. As 
the system evolves these quantities grow monotonically, but while at 
$t=20$ for the velocity field $\ell \approx 0.5$, for the magnetic field 
$\ell \approx 1$. As a criterion to decide if the simulations were well 
resolved in spectral space, the scales where the kinetic and magnetic 
enstrophy spectrum peaked were observed as a function of time, and it 
was asked that their corresponding wavenumbers $|\lambda|$ were smaller 
than $\max\{|\lambda|\}$ at all times.

Figure \ref{fig:decay_xi} shows the amplitude of the individual 
modes $\xi^B$ in run II at $t=10$ as a function of $q$ and $l$ (all 
values of $m$ for each value of $l$ are summed). Most of the magnetic 
energy is concentrated in the shell with $l=1$, and the modes with 
$q = \pm 1$ in this shell have the largest amplitude. Note the 
imbalance between the mode with $q=1$ and $q=-1$, indicating some 
helicity is present in the magnetic field.

The dominance of a helical large scale magnetic field at late 
times can also be identified in an inspection of the fields in 
configuration space. Figure \ref{fig:decay3D} shows field intensity 
and field lines for the velocity and magnetic field at $t=1.5$ 
(left) and $t=15$ (right) in run II. While at early times both fields 
show small scale features, at late times the velocity field looks 
reminiscent of a quadrupole and the magnetic field looks like a 
dipole oriented roughly in the $z$ direction. However, the magnetic 
field at $t=15$ is helical and the magnetic field lines are not 
purely poloidal. There is a small toroidal component to the magnetic 
field, and the magnetic field lines proceed slowly in the 
$\phi$-direction in a helical fashion.

In Fig. \ref{fig:decay3D} and in the following visualizations 
the labels are as follows. The $x$, $y$, and $z$ directions are 
indicated by the arrows (in the online version, these are respectively 
red, green, and blue). \ad{The color and opacity are proportional to the 
field intensity, colorbars are given as a reference.} Field lines are 
computed taking a snapshot of the field at a fixed instant in time, 
\ad{and integrating a trajectory from twenty random points} in the 
surroundings of the center of the sphere. The field is not evolved 
in time as the lines are integrated. To indicate the direction of the 
fields, in the online version the lines change color according to the 
distance integrated from the initial point; from \ad{red to yellow, and 
finally blue.}

\section{\label{sec:application2}DYNAMO EFFECT}

\begin{figure}
\includegraphics[width=7cm]{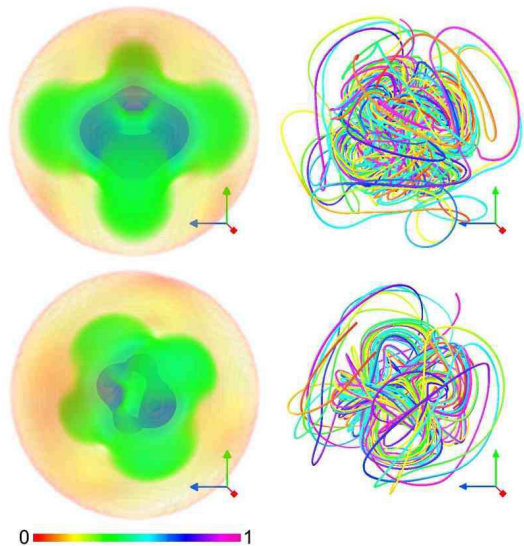}
\caption{Intensity of the forcing function ${\bf f}$ used in the dynamo 
    simulations (left), and associated field lines (right). Above: 
    function used in run III. Below: function used in runs IV and V.}
\label{fig:force}
\end{figure}

\begin{figure}
\includegraphics[width=9cm]{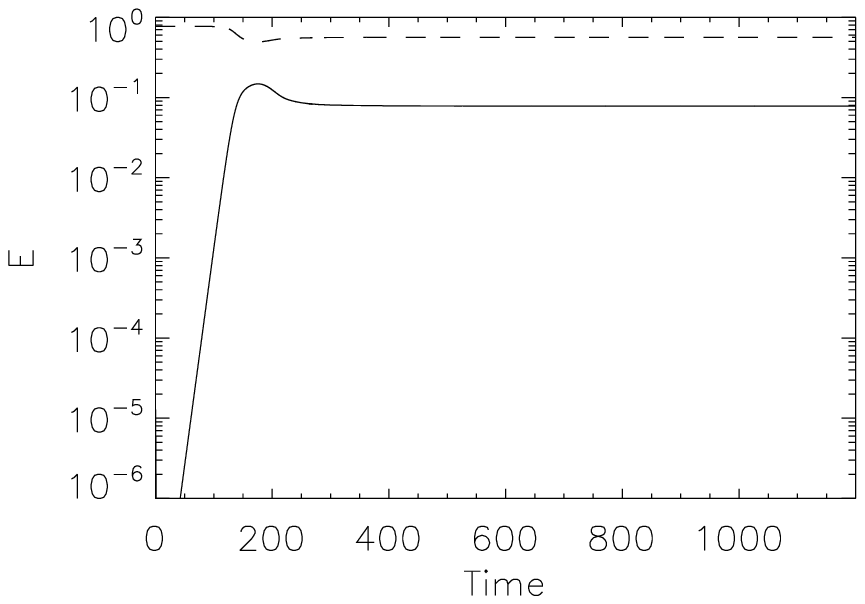}
\caption{Kinetic (dashed line) and magnetic energy (solid line) as a 
    function of time in run III.}
\label{fig:dynamo_ener1}
\end{figure}

\begin{figure}
\includegraphics[width=9cm]{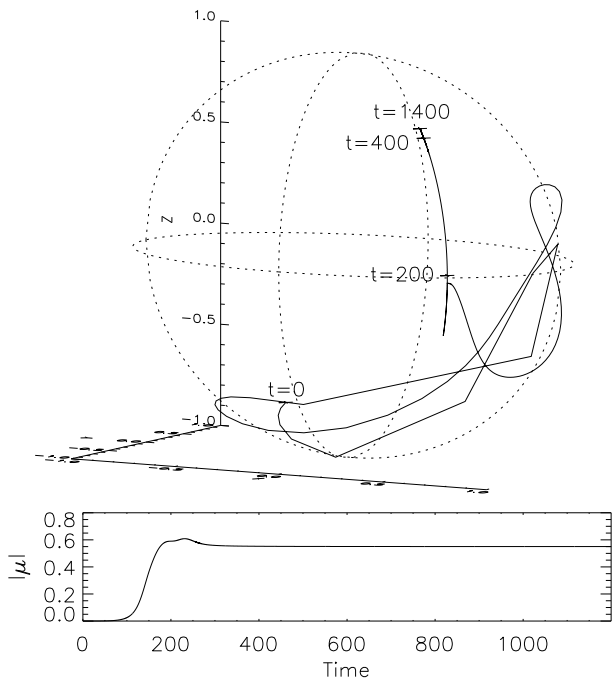}
\caption{\ad{Trace of the magnetic dipole orientation on the unit sphere 
    (above) and magnitude of the dipole moment as a function of time 
    (below) in run III.}}
\label{fig:dynamo_dip1}
\end{figure}

\add{In an MHD dynamo, an initially small ``seed'' magnetic field is 
amplified and sustained against Ohmic dissipation solely by the motions 
of a conducting fluid. Magnetic fields observed in planets and stars 
are believed to be the result of a dynamo process. The mechanical 
mechanisms proposed vary widely, including thermal convection, Ekman 
pumping due to rotation, precession, irregularities on the inner 
surface of the planetary mantles, and so on. In this Section, we 
study three simple examples of forced dynamo action in the sphere 
with relatively simple realizations of the forcing function ${\bf f}$ 
of Eq. (5).} \ad{Figure \ref{fig:force} shows visualizations of the two 
expressions used for ${\bf f}$. The geometry of the forcing function is 
not intended to mimic any particular process in the planetary or stellar 
dynamos, but is rather inspired in a common practice in dynamo simulations 
with periodic boundary conditions: a few spectral modes are forced in the 
mechanical energy, and for Reynolds numbers large enough generic 
properties of the simulations are studied.}

\add{In all cases to be described here, a purely hydrodynamic 
(${\bf B}=0$) computation was carried out first until a steady state 
(laminar or turbulent) was reached for the velocity field. The 
amplitude of ${\bf f}$ was chosen to have rms velocity of order 
one in the steady state. Then a random magnetic field with energy 
$E_M \sim 10^{-6}$ was loaded into the modes with $|q|=l=4$.} The 
simulations were then continued in order to observe the amplification 
and subsequent development of the magnetic field.

Three dynamo simulations were done in which the Reynolds numbers and 
the number of modes excited were progressively increased. \add{In the 
first case, the system reaches a steady state with a stable dipole 
moment which varies little in time. In the second case, the dipole 
moment forms, then spontaneously changes direction. In the third, the 
Reynolds number is high enough that the velocity field might be called 
turbulent, and many modes are excited; for this case, the dipole moment 
changes erratically in time.}

A resolution of $\max \{|q|\} = \max \{l\} = 9$ was used in all the 
runs. The same criteria as in the previous section was used to decide 
if a simulation was well resolved. For the last run, a simulation with 
higher resolution was also carried out to see if the results would be 
modified by the change in resolution.

\begin{figure}
\includegraphics[width=7cm]{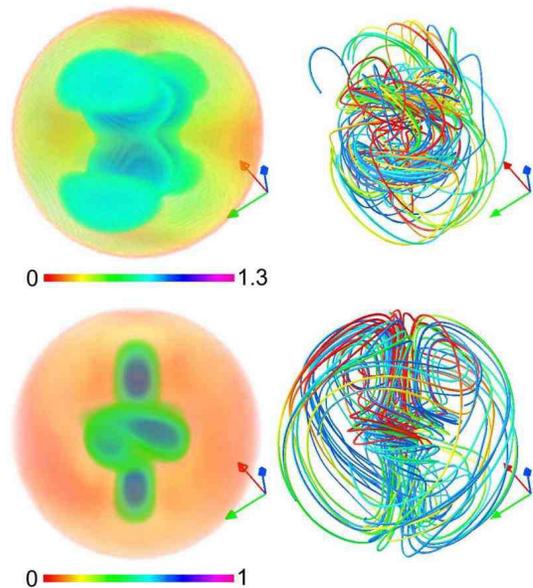}
\caption{Above: kinetic energy density and velocity field lines in run 
    III, at $t=915$. Below: magnetic energy density and magnetic field 
    at the same time. \ad{The same conventions than in Fig. \ref{fig:decay3D} 
    are used.}}
\label{fig:3Dlaminar}
\end{figure}

\subsection{\label{sec:laminar}Laminar runs}

In this section we present results from two runs with 
$\nu=\eta=3\times10^{-3}$. In run III, the external forcing ${\bf f}$ 
in Eq. (\ref{eq:momentum}) is given by the coefficient
\begin{equation}
\xi^f_{2,2,1} = f_0(1+i),
\end{equation}
with $f_0 = 1.4$, which corresponds to one C-K mode and as a result 
injects maximum kinetic helicity in the system. In run IV, the external 
forcing is
\begin{eqnarray}
& \xi^f_{2,2,0} = 5 \xi^f_{-2,2,0} = f_0 , 
    \label{eq:force1} \\
& \xi^f_{2,2,m} = 5 \xi^f_{-2,2,m} = f_0(1+i) ,
    \label{eq:force2}
\end{eqnarray}
where $m$ runs from 1 to 2 and negative values of $m$ are again given 
by $\xi_{q,l,-m} = (-1)^m \xi^*_{qlm}$. The amplitude of the forcing 
is $f_0 = 0.9$. This forcing injects non-maximal kinetic helicity (plus, 
of course, kinetic energy) into the system. \ad{Figure \ref{fig:force} shows 
visualizations of the two forcing functions in configuration space. In 
both cases the forcing is stronger in the center of the sphere, and a 
modulation due to $m=2$ modes in the forcing can be easily identified.}
The phase and amplitude of the external force ${\bf f}$ were kept 
constant during the entire simulations.

Figure \ref{fig:dynamo_ener1} shows the time history of the 
magnetic and kinetic energy in run III. The Reynolds numbers based on 
the length $R=1$ and the rms velocity are $R_V = R_M \approx 290$, and 
the energy containing scale of the flow is $\ell \approx 0.5$. Before 
the magnetic field is introduced, only the forced mode is excited.

After the magnetic seed is introduced, the magnetic energy is 
amplified exponentially in a kinematic regime. Then the magnetic field 
saturates around $t\approx 150$ and the Lorentz force quenches the 
velocity field. In the final steady state, more mechanical modes 
besides the forced mode are excited. This is the result of an 
instability of the flow, triggered by the Lorentz force as the 
magnetic field grows exponentially.

Figure \ref{fig:dynamo_dip1} shows \ad{the trace of the orientation of 
the dipole moment in the surface of the unit sphere, and the magnitude 
of the dipole moment as a function of time for run III.} The dipole 
moment grows during the kinematic regime, \ad{but its orientation changes 
erratically in time. At $t \approx 200$ $|\vmu|$ reaches its maximum 
amplitude and converges slowly to a steady state. Also it direction 
changes slowly, and after $t\approx 400$ almost no change is observed. 
At late times} the dipole shows no inclination toward further systematic 
dynamical development.

\begin{figure}
\includegraphics[width=9cm]{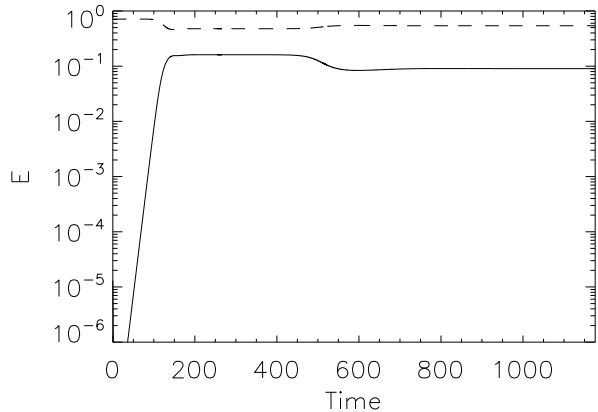}
\caption{Kinetic (dashed line) and magnetic energy (solid line) as a 
    function of time in run IV.}
\label{fig:dynamo_ener2}
\end{figure}

\begin{figure}
\includegraphics[width=9cm]{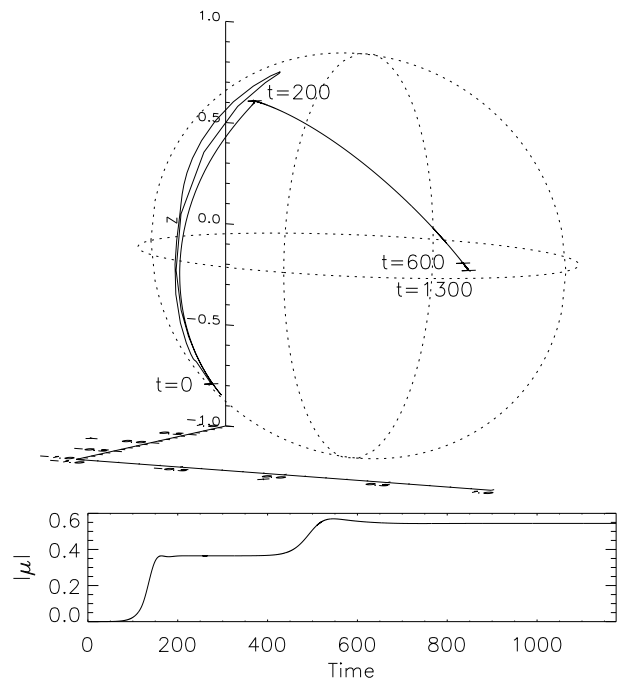}
\caption{\ad{Trace of the magnetic dipole orientation on the unit sphere 
    (above) and magnitude of the dipole moment as a function of time 
    (below) in run IV.}}
\label{fig:dynamo_dip2}
\end{figure}

A visualization of the velocity and magnetic fields in configuration 
space in the steady state of run III is shown in Fig. \ref{fig:3Dlaminar}. 
The kinetic energy is concentrated in two counter-rotating regions, located 
in the center of each hemisphere. \ad{Note the $m=2$ modulation in the 
forcing is still visible in the kinetic energy density.} The velocity 
field in these regions is mostly toroidal, as indicated by the velocity 
field lines. The magnetic energy is larger in the center of the sphere, 
and along the axis defined by the two counter-rotating eddies. In the 
interior, but away from the axis, magnetic field lines are mostly 
toroidal, as the result of the stretching by the two counter-rotating 
eddies. Along the axis and close to the boundaries, the flow is mostly 
poloidal.

\begin{figure}
\includegraphics[width=9cm]{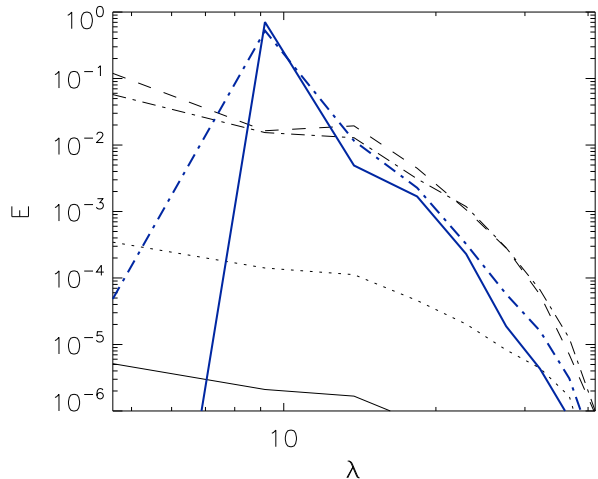}
\caption{Kinetic energy spectrum at $t=52$ [thick (blue) solid line] and 
    at $t=500$ [thick (blue) dash-dotted line] in run IV; the thin lines 
    correspond to the magnetic energy spectrum at $t=52$ (solid), $t=82$ 
    (dotted), $t=250$ (dashed), and $t=500$ (dash-dotted).}
\label{fig:dynamo_spc2}
\end{figure}

\begin{figure*}
\includegraphics[width=14cm]{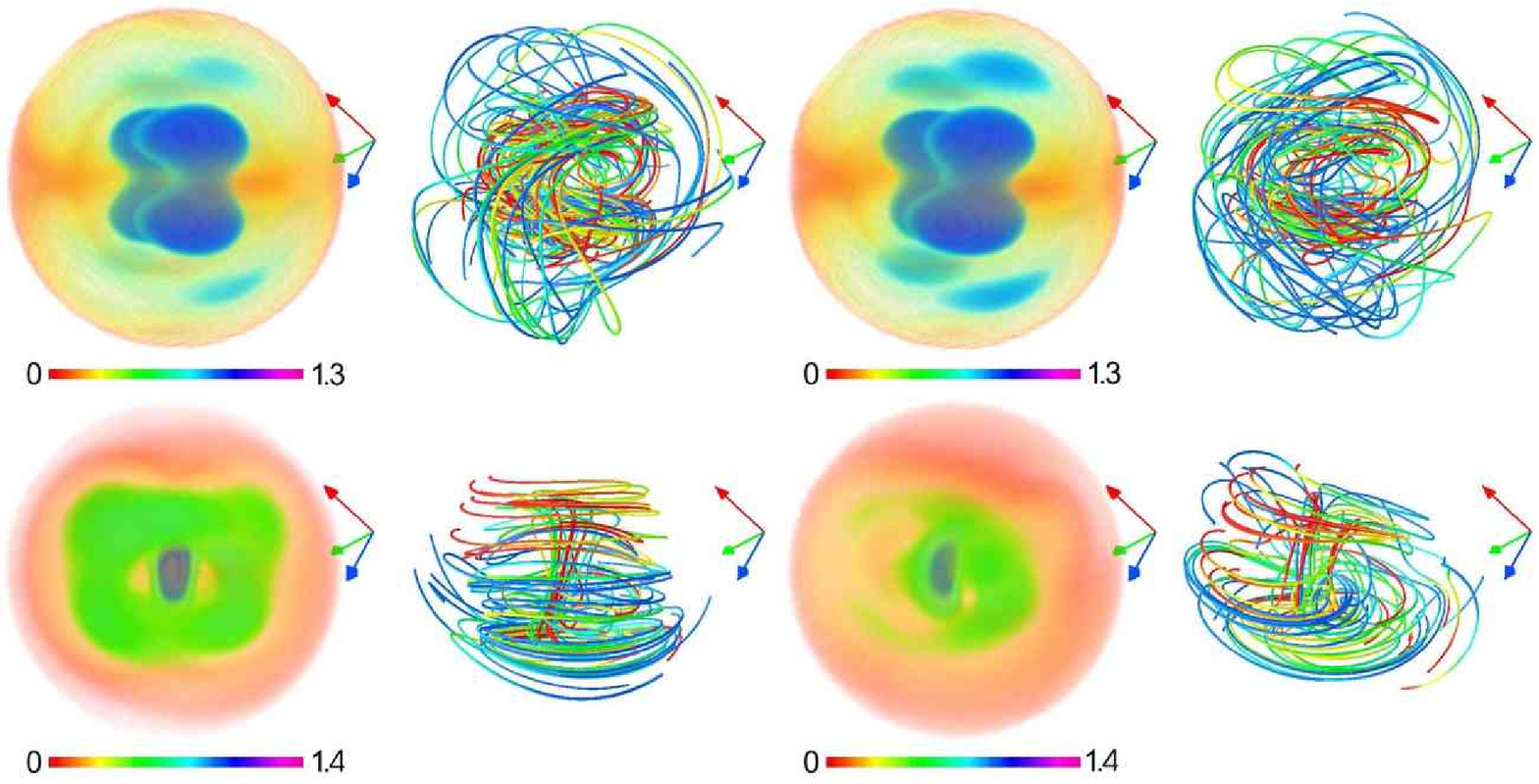}
\caption{Above: kinetic energy density and velocity field lines in run 
    IV, at $t=315$ (left) and at $t=1065$ (right). Below: magnetic energy 
    density and magnetic field lines in the same run, at $t=315$ (left) 
    and at $t=1065$ (right). \ad{The same conventions than in Fig. 
    \ref{fig:decay3D} are used.}}
\label{fig:reversal}
\end{figure*}

The time history of the kinetic and magnetic energy in run IV 
is shown in Fig. \ref{fig:dynamo_ener2}. The Reynolds numbers for this 
run (based on the length $R=1$) are $R_V = R_M \approx 280$, and the 
energy containing scale of the flow is $\ell \approx 0.48$. Although 
the kinematic viscosity and magnetic diffusivity are the same as in 
run III, the external forcing injects energy in a larger number of 
modes and even before the magnetic field is introduced non-forced modes 
have some mechanical energy. After the magnetic seed is introduced, the 
magnetic energy grows exponentially until at $t \approx 150$ saturates. 
The system seems to reach a steady state but suddenly at $t\approx 500$ 
the magnetic energy decreases by a factor of $\approx 1.8$, the kinetic 
energy increases by $\approx 1.1$, and the system reaches a second 
steady state.

\begin{figure}
\includegraphics[width=9cm]{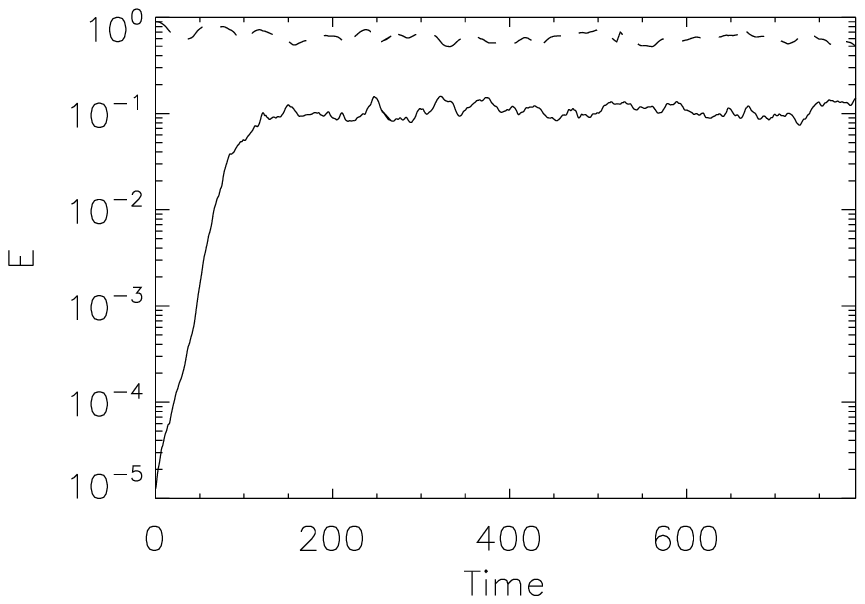}
\caption{Kinetic (dashed line) and magnetic energy (solid line) as a 
    function of time in run V.}
\label{fig:dynamo_ener3}
\end{figure}

\begin{figure}
\includegraphics[width=9cm]{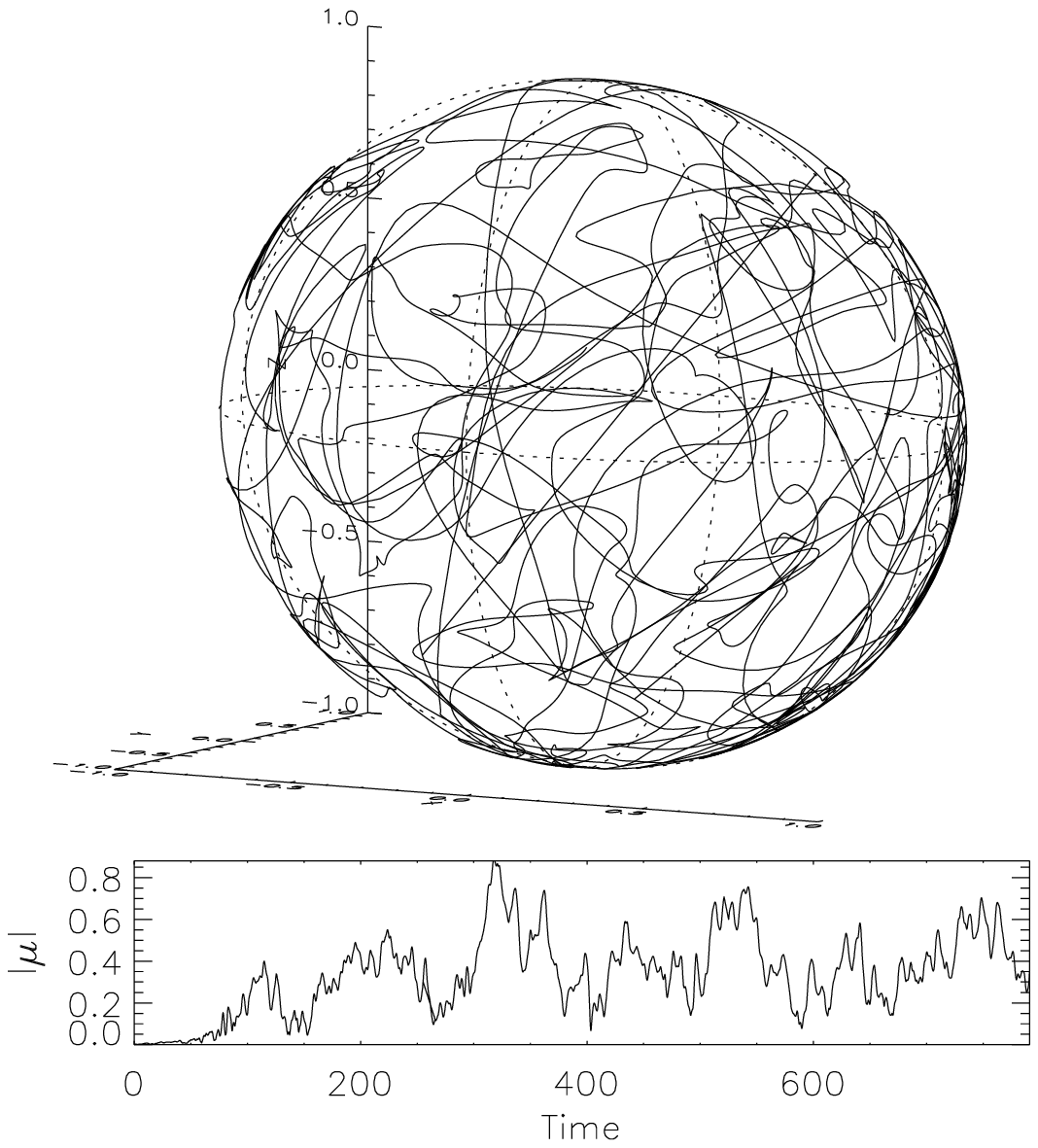}
\caption{Time evolution of the three components of the dipole moment 
    in run V. Labels are as in Fig. \ref{fig:dynamo_dip1}.}
\label{fig:dynamo_dip3}
\end{figure}

The abrupt change in the kinetic and magnetic energy in run IV at 
$t \approx 500$ is associated with a reorientation of the magnetic dipole 
moment. Figure \ref{fig:dynamo_dip2} shows the time evolution \ad{of the 
direction and amplitude of $\vmu$. As in run III, in the kinematic stage 
the dipole moment grows rapidly and its direction fluctuates erratically, 
until reaching a first quasi-steady state at $t \approx 200$. The dipole 
moment then stays approximately constant until at $t \approx 400$} the 
magnetic field evolves rapidly, and the dipole moment changes direction 
to a second attractor \ad{(reached at $t \approx 600$)}. The angle the 
dipole flips by is close to $\pi/2$. \ad{After $t \approx 600$ only a slow 
change in $\vmu$ is observed.} The amplitude of the dipole moment 
also changes rapidly as the dipole shifts at $t \approx 400$; while at 
$t \approx 300$ $|\vmu| \approx 0.36$, at 
$t \approx 1000$ $|\vmu| \approx 0.54$.

\begin{figure}
\includegraphics[width=9cm]{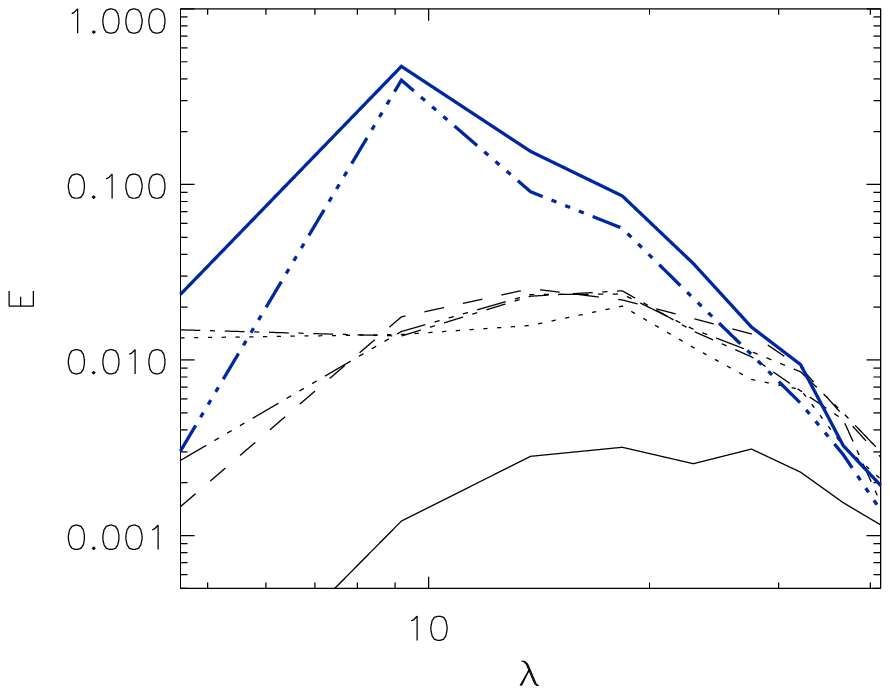}
\caption{Kinetic energy spectrum at $t=75$ [thick (blue) solid line] and 
    at $t=600$ [thick (blue) dash-triple dotted line] in run V; the thin 
    lines correspond to the magnetic energy spectrum at $t=75$ (solid), 
    $t=168$ (dotted), $t=240$ (dashed), $t=544$ (dash-dotted), and $t=500$ 
    (dash-triple dotted).}
\label{fig:dynamo_spc3}
\end{figure}

Figure \ref{fig:dynamo_spc2} shows the kinetic and magnetic energy 
spectra in run IV at different times. The kinetic energy spectrum peaks 
at $|\lambda| \approx 9$, corresponding to the forced modes with $q=2$ 
and $l=2$. The energy in the remaining modes is at least two orders
of magnitude smaller than in the forced modes. At late times, some 
kinetic energy is excited in the largest available scale, as well as a 
small angular momentum ($|{\bf L}|^2/E_V \approx 1.5 \times 10^{-4}$ 
after $t \approx 600$). The angle between the dipole moment and this 
small angular momentum remains constant after $t=100$ and is $\pi/2$. 
Visualizations of the velocity and magnetic fields in configuration 
space are shown in Fig. \ref{fig:reversal}. The geometry of the 
velocity field is more complex than in run III\ad{, although the $m=2$ 
modulation in the kinetic energy can still be recognized.} Note at 
$t=315$ the magnetic field lines are mostly poloidal in the center 
of the sphere, and toroidal close to the boundary. \ad{The change in 
the orientation of $\vmu$ at $t \approx 400$ takes place without a 
strong change in the velocity field, and a relatively small change in 
the magnetic field configuration.}

\begin{figure*}
\includegraphics[width=14cm]{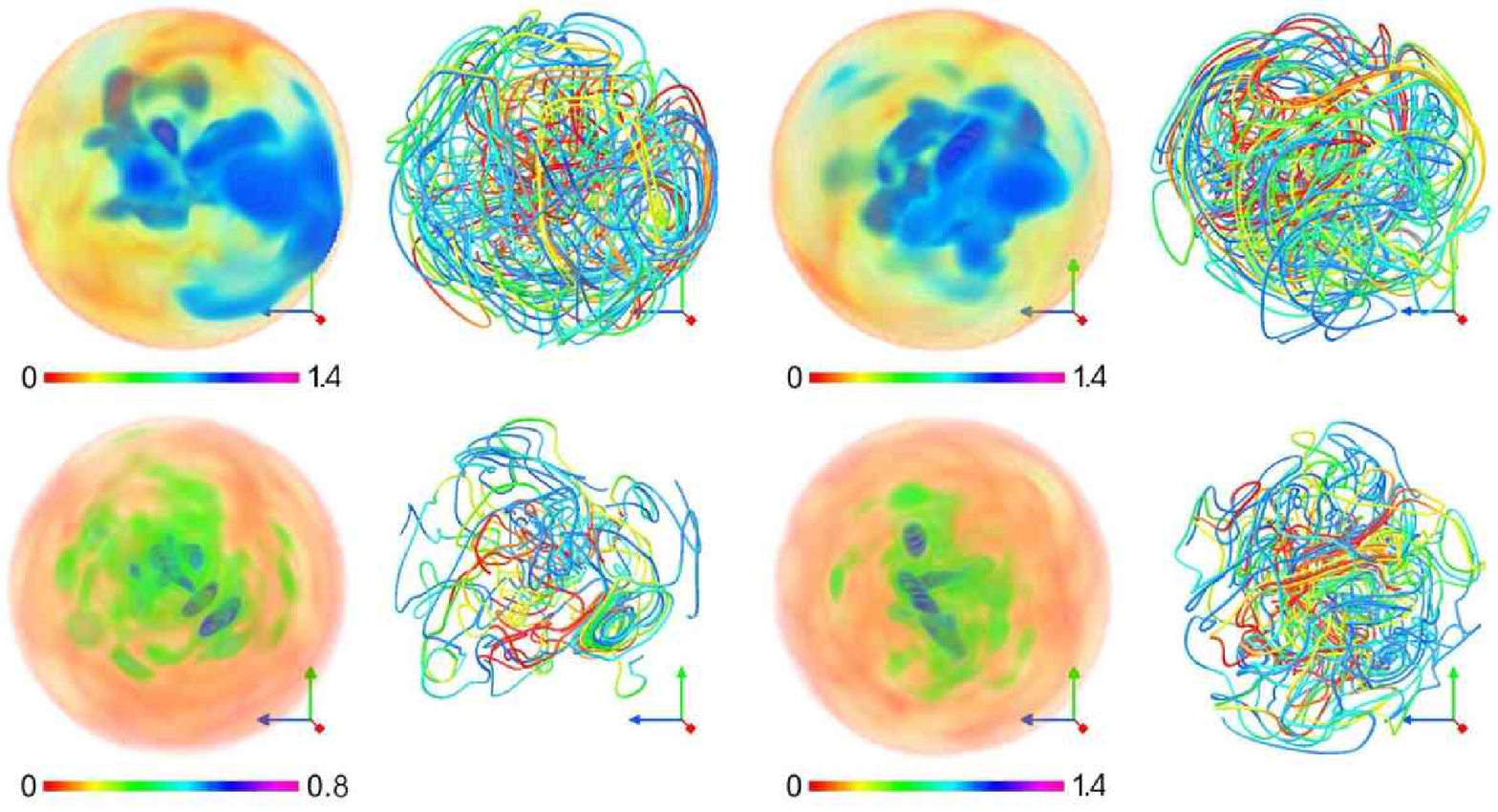}
\caption{Above: kinetic energy density and velocity field lines in run 
    V, at $t=75$ (left) and at $t=1035$ (right). Below: magnetic energy 
    density and magnetic field lines in the same run, at $t=75$ (left) 
    and at $t=1035$ (right). \add{We were unable to find an ordered 
    pattern in either field.}}
\label{fig:dynamo3D}
\end{figure*}

\subsection{\label{sec:chaos}A \add{rapidly-varying dipole}}

For run V, the external forcing f is given by Eqs. (\ref{eq:force1}) and 
(\ref{eq:force2}), but the kinematic viscosity and magnetic diffusivity 
are dropped to $\nu = \eta = 3 \times 10^{-4}$. The resulting kinetic 
and magnetic Reynolds numbers based on the length $R=1$ are 
$R_V=R_M \approx 2300$, and the energy-containing and Taylor scales are 
respectively $\ell \approx 0.4$ and $\lambda_T = 0.35$.

The evolution of the kinetic and magnetic energy in run V is shown in 
Fig. \ref{fig:dynamo_ener3}. Again, after an initial kinematic 
regime where the magnetic energy is amplified exponentially, the system 
reaches a statistical steady state. Note that in this simulation both 
the kinetic and magnetic energy fluctuate strongly with time, indicating 
the nonlinear coupling between modes is stronger than in runs III 
and IV, as a result of the higher Reynolds numbers.

\ad{Figure \ref{fig:dynamo_dip3} shows the trace of the dipole moment in 
the surface of the unit sphere, and its intensity as a function of time, 
in run V. The direction of $\vmu$ appear to fluctuate randomly, changing 
hemisphere with a characteristic time of the order of 100 eddy turnover 
times. In the meantime, the whole surface of the unit sphere seems to be 
explored by the fluctuations in $\vmu$.} The angular momentum is small 
and fluctuates around $|{\bf L}|^2/E_V \approx 5 \times 10^{-3}$. However, 
unlike in Run III, in this simulation the angle between the dipole moment 
and the angular momentum is not even approximately constant and fluctuates 
seemingly randomly between $0$ and $\pi$. The question of under what 
circumstance the dipole favors one or another alignment appears to be 
quite unsettled, and deserves further study in higher-resolution 
simulations with larger angular momentum \ad{and rotation}.

The kinetic and magnetic energy spectra in run V at several 
times are shown in Fig. \ref{fig:dynamo_spc3}. More modes are excited 
in the velocity field, in accordance with the strong fluctuations in 
time observed in the kinetic energy. While in runs III and IV the 
magnetic energy spectrum peaks at large scales even during the kinematic 
regime, in run V at early times the magnetic energy peaks at scales 
smaller than the forcing scale. Also, after the nonlinear saturation of 
the dynamo, a fluctuation in the amplitude of the large scale magnetic 
field is observed. The minima are correlated with times of minima 
of $\mu^2$, when the three components of the dipole moment fluctuate 
around zero. Most of the activity in this run is in small scales 
and fluctuations in the flow are larger than in runs III and IV. Even 
at late times when a large scale magnetic field has developed, 
intermediate scales give a large contribution to the magnetic energy. 
This also explains the strong fluctuations observed in the dipole 
moment. Since $\vmu$ is proportional to the current density, the 
small scales give a large contribution to the dipole moment.

Figure \ref{fig:dynamo3D} shows the magnetic and velocity field 
in real space at two different times in run V. The fields have more 
small scale structure than in run IV. \ad{Any trace of the $m=2$ 
modulation of the forcing has been completely lost.} Also, in the 
kinematic regime (see e.g. the magnetic field at $t=75$) the magnetic 
energy is mostly in the small scales, as also indicated by the 
magnetic energy spectrum.

\section{\label{sec:conclusions}DISCUSSION}

\add{We have introduced some computational machinery that is intended for 
a quantitative discussion of nonlinear and incompressible 
magnetohydrodynamics inside a sphere for a wide range of initial 
conditions and varieties of mechanical forcing. It is concerned 
essentially with the analytical and computational aspects of 
dynamo action in this geometry in a somewhat abstract setting, and 
is not the same as periodic dynamo computations with rectangular 
symmetry, or with planetary-dynamo or solar-dynamo computations 
whose central focus is reproduction of observations of magnetic 
behavior of a real system. There are other features yet to be 
included and it is our intent to introduce them one at a time: rigid 
rotation and Ekman pumping, an insulating but non-conducting boundary 
to permit the magnetic field's penetration into the surrounding vacuum 
region, and different forms of mechanical excitations.}

\add{The operation of the wholly spectral code, which has had a precedent 
in axially periodic circular-cylinder geometry has been illustrated by a 
few simple examples, not in any sense a comprehensive study. First, 
decaying turbulence has been studied involving the relaxation of helical 
initial conditions to a magnetically dominated state whose spectrum is 
dominated by the longest allowed spatial scales and whose kinetic energy 
is essentially gone. Also, dynamo computations have been done for three 
successively higher sets of Reynolds numbers, revealing a different magnetic 
behavior in each case. In the first case, a magnetic dipole formed in what 
was essentially a laminar velocity field and appears willing to persist for 
as long as the code is run. In the second case, another dipole formed, but 
after the passage of a few hundred large-scale eddy-turnover times, it 
changed its magnitude to a somewhat larger value, and flipped its 
orientation, for reasons we do not know but that are worth exploring. 
These fluctuations have required neither a preferred direction enforced 
by rigid rotation nor thermally convective rolls. Finally, in the third 
case, the velocity field had Reynolds numbers (based on the radius of 
the sphere) of about $2300$ and could be said to be turbulent. In this 
turbulent case, there was a dipole moment, but it exhibited no systematic 
or regular behavior as far as we could tell, and changed its orientation} 
\ad{from one hemisphere to the other} \add{every $100$ or so eddy turnover 
times,} \ad{exploring in the meantime all possible orientations.} \add{It 
did appear to be a resolved computation, despite the turbulence, and we 
believe represents a bona fide solution of the MHD equations. All these 
runs, it should be stressed, were carried out with a magnetic Prandtl 
number of unity, and may well change for values far from that. One 
outcome to be noted is that neither turbulence nor rotation have been 
a necessary ingredient for the development and maintenance of a 
magnetic dipole, but the presence of mechanically helicity has 
helped a lot. Spontaneous changes in magnetic dipole orientation 
have been easy to observe, both in laminar and turbulent cases.}

\add{We should also stress that the code as presently constituted is 
limited to a relatively small number of degrees of freedom, far 
fewer than well-resolved high Reynolds number mechanical turbulence 
would demand. It will also be attempted to design wall-friction 
terms to permit a more efficient exchange of angular momentum from 
the fluid to the wall \cite{Shan94}, once rigid rotation is introduced.} 
\ad{Also rigid rotation and different boundary conditions for the magnetic 
fields can be implemented. Given the properties and limitations of 
the method described (purely spectral, non-dispersive, and 
conservative), we believe this method can be used as a testbed to 
explore the effect of different physical effects, boundary conditions, 
subgrid models (several models, such as the Lagrangian averaged model 
\cite{Holm02a,Holm02b,Mininni05d} are easier to implement in spectral 
space), etc., before trying these ideas in more complex and realistic 
codes to reach high Reynolds numbers.}

\begin{acknowledgments}
The authors would like to express their gratitude to A. Pouquet for 
valuable discussions and his careful reading of the manuscript. Computer 
time was provided by NCAR. The NSF grants CMG-0327888 at NCAR and 
ATM-0327533 at Dartmouth College supported this work in part and are 
gratefully acknowledged. Three-dimensional visualizations of the flow 
were done using VAPoR \cite{vapor}, a software for interactive 
visualization and analysis of large datasets.
\end{acknowledgments}

% BIBLIOGRAPHY %%%%%%%%%%%%%%%%%%%%%%%%%%%%%%%%%%%%%%%%%%%%%%%%%%%%%%%

%\bibliography{ms}

\end{document}